\documentclass[conference]{IEEEtran}
\IEEEoverridecommandlockouts

\usepackage{cite}
\usepackage{amsmath,amssymb,amsfonts}
\usepackage{algorithmic}
\usepackage{graphicx}
\usepackage{textcomp}
\usepackage{xcolor}
\usepackage{bm}
\def\BibTeX{{\rm B\kern-.05em{\sc i\kern-.025em b}\kern-.08em
    T\kern-.1667em\lower.7ex\hbox{E}\kern-.125emX}}

\makeatletter 
\newcommand{\linebreakand}{%
  \end{@IEEEauthorhalign}
  \hfill\mbox{}\par
  \mbox{}\hfill\begin{@IEEEauthorhalign}
}
\makeatother 

\begin{document}

\title{$X$-$Z$ Round Scheduling for the Surface Code with Defects under Biased Noise}


\author{
    \IEEEauthorblockN{
        Lakshika Rathi\IEEEauthorrefmark{1},
    Pau Escofet\IEEEauthorrefmark{1}\IEEEauthorrefmark{2},
    Joshua Viszlai\IEEEauthorrefmark{3},
    Margaret Martonosi\IEEEauthorrefmark{1}\\
    }
    \IEEEauthorblockA{
        \textit{\IEEEauthorrefmark{1}Princeton University, USA}\\
        \textit{\IEEEauthorrefmark{2}Universitat Politècnica de Catalunya, Spain}\\
        \textit{\IEEEauthorrefmark{3}University of Chicago, USA}
    }
    \{lakshika, mrm\}@princeton.edu, pau.escofet@upc.edu, viszlai@uchicago.edu
}

\newcommand{\LR}[1]{\textcolor{blue}{[#1]\textsubscript{LR}}}
\newcommand{\PE}[1]{\textcolor{red}{[#1]\textsubscript{PE}}}

\maketitle

\IEEEpubid{\begin{minipage}{\textwidth}\ \\[45pt]
\centering Accepted at the 2026 IEEE International Conference on Quantum Computing and Engineering (QCE)
\end{minipage}}

\begin{abstract}

Fault-tolerant Quantum Computing (FTQC) relies on Quantum Error Correction (QEC) codes that encode logical qubits across many physical qubits to detect and correct errors. The surface code is among the most widely studied codes due to its high error threshold, the existence of efficient decoders, and hardware-friendly properties: a planar, two-dimensional layout with nearest-neighbor connectivity. In practice, however, the fabrication of solid-state quantum processors introduces hardware defects, resulting in defective qubits and couplers that must be discarded. Adapting the surface code to these defects often requires measuring the $X$- and $Z$-type checks in separate rounds rather than simultaneously.
In this work, we investigate the optimal $X$-to-$Z$ checks round-scheduling ratio under biased noise systems. Our results characterize how key architectural parameters, such as noise bias, code distance, and defect rate, impact the logical error rate. We provide insights into how to determine the optimal scheduling ratio directly from device calibration data, enabling manufacturers to maximize performance without extensive simulations. Our approach reduces the logical error rate by up to $4.25\times$ at a $1\%$ defect rate and up to $8.46\times$ at a $2\%$ defect rate for a distance-$13$ surface code under moderately biased noise. Furthermore, we demonstrate that the benefits of round-scheduling extend beyond the defective-hardware setting. In biased-noise architectures subject to CNOT crosstalk, separating $X$ and $Z$ measurement rounds yields up to $4.5\times$ reduction in logical error rate.



\end{abstract}


\section{Introduction}
\label{sec:introduction}

Fault-tolerant Quantum Computing (FTQC) requires Quantum Error Correction (QEC) to protect encoded logical information from errors in the underlying physical qubits \cite{preskill1997faulttolerantquantumcomputation}. The surface code \cite{Fowler_2012} is a leading candidate for near-term implementation, owing to its high threshold, compatibility with nearest-neighbor connectivity, and the availability of fast, scalable decoders \cite{Higgott2025sparseblossom, Wu_2025}. Recent experiments on superconducting platforms have already demonstrated surface code memory below the physical error rate threshold \cite{2024}, bringing practical fault tolerance within reach. 

In practice, the fabrication of solid-state quantum processors inevitably yields a fraction of inoperable qubits and couplers, arising from either material defects \cite{Bilmes_2020} or two-level systems \cite{Klimov_2018, M_ller_2019}. Such components cannot be used for computation and need to be excluded from the processor. Adapting the surface code to these defects without sacrificing its fault-tolerant properties is an active area of research \cite{Leroux_2025, Debroy_2025, wolanski2026automatedcompilationincludingdropouts}. A feature of the resulting schemes is that the $X$- and $Z$-type checks near a defect can no longer be measured simultaneously and are instead measured in alternating rounds \cite{Leroux_2025}.  

\begin{figure*}
    \centering
    \includegraphics[width=1\linewidth]{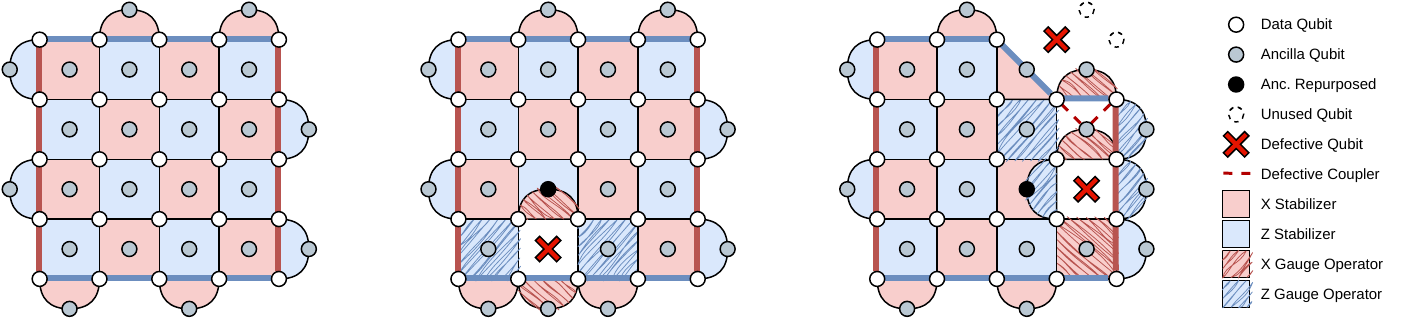}
    \caption{\textbf{(left)} Distance-5 rotated surface code with $d^2$ data qubits and $d^2-1$ ancilla qubits which measure $X$- and $Z$-type stabilizers. \textbf{(middle)} Surface code under the presence of 1\% defect rate, with a defective ancilla qubit, necessitating two super-stabilizers (one $X$, one $Z$) to obtain the syndrome. \textbf{(right)} Surface code under the presence of 2\% defect rate, with one defective ancilla, one defective data qubit, and two defective couplers.}
    \label{fig:surface_code_defects}
\end{figure*}

Additionally, physical qubits on most leading platforms exhibit asymmetric Pauli noise channels, where dephasing errors dominate over bit-flips \cite{Aliferis_2008, Qing:2024xfv}. Both of these hardware challenges have received attention individually. To address the noise asymmetry, prior work modifies the code itself: the XZZX surface code \cite{Bonilla_Ataides_2021} and tailored CSS variants \cite{Tuckett_2018} both exploit bias at the stabilizer level to achieve higher thresholds. 

To our knowledge, however, the joint setting (i.e., a defect-adapted surface code operating under biased noise) has not been systematically studied. 
Defect adaptation methods \cite{Leroux_2025} convert the affected stabilizer into lower-weight gauge checks that anti-commute with one another, requiring an alternating $X$/$Z$ syndrome extraction schedule. This schedule is typically chosen uniformly, with an equal number of rounds in each basis. Under biased noise, however, a uniform schedule is suboptimal. For instance, under $Z$-biased noise, dephasing errors on data qubits dominate over bit-flips; allocating more rounds to $X$-type checks to sample the dominant-error syndromes more often, reduces the logical error rate (LER). 

In this work, we systematically investigate the optimal $X$-to-$Z$ round scheduling ratio for defect-adapted surface codes in the low-to-moderate bias regime. Through extensive numerical simulations, we characterize how a well-chosen ratio depends on three key architectural parameters: noise bias ($\eta = p_z/p_x$), code distance, and defect rate. We show that this scheduling yields substantial reductions in LER relative to the uniform $1{:}1$ baseline. We provide insights into how to choose the ratio directly from measured calibration data, eliminating the need for per-device simulation sweeps. Finally, we demonstrate that the benefits of scheduling extend beyond the defective setting. In quantum systems with crosstalk errors under biased noise, separating $X$ and $Z$ basis syndrome extraction rounds reduces the LER. 

Our main contributions are: 
\begin{itemize}
    \item We establish that under biased noise, round scheduling is a first-order design parameter for defect-adapted surface codes. The optimal scheduling ratio departs from $1{:}1$. Choosing it correctly yields a $1.12\times$-$8.46\times$ improvement in the logical error rate across the considered bias regimes ($\eta \leq 10$).
    \item We characterize how the optimal ratio depends on noise bias, code distance, and defect rate. In particular, we find it is set primarily by the noise bias and is, to a good approximation, independent of the code distance. This allows system designers to select the ratio directly from measured calibration data, without running device-specific simulations. 
    \item We show that the benefits of round scheduling generalize beyond defects. For example, consider crosstalk-dominated hardware under biased noise, these approaches improve the average logical error rate by up to $4.5\times$ for a distance-$13$ surface code when parallel CNOTs incur a $2\times$ crosstalk penalty. 
\end{itemize}

The remainder of this paper is organized as follows. Section \ref{sec:background} reviews the surface code, how to adapt it to a defective architecture, and biased-noise systems. Sections \ref{sec:methodology} and \ref{sec:results} describe our methodology and the results obtained on round scheduling under several biased noise settings. Section \ref{sec:crosstalk} extends the analysis to crosstalk-dominated architectures. Section \ref{sec:related work} discusses related work. Lastly, we conclude the paper and discuss future directions of this work in Section \ref{sec:conclusion}. 

\section{Background}
\label{sec:background}

\subsection{Quantum Error Correction and the Surface Code}

QEC codes are essential for large-scale FTQC \cite{2024, Preskill_2018}. They encode a single logical qubit into many physical qubits to protect quantum information against decoherence, bit flips, phase flips, and other error mechanisms \cite{Nielsen_Chuang_2010}. Provided the physical error rate is below a certain fault-tolerance threshold, the LER can be suppressed exponentially by increasing the code distance ($d$). 

The surface code \cite{Fowler_2012} uses $d^2$ physical qubits to encode a single logical qubit and can correct up to $\lfloor(d-1)/2\rfloor$ physical errors. It is one of the leading candidates for QEC due to its nearest-neighbor connectivity requirements, which make it suitable for solid-state hardware \cite{Nakamura_1999, 48651}; its high physical noise threshold, which has already been reached in current processors \cite{2024}; and the existence of fast, reliable decoding algorithms, such as minimum-weight perfect matching (MWPM) \cite{Higgott2025sparseblossom}. The surface code with distance $d=5$ is depicted in Figure~\ref{fig:surface_code_defects} left.

In the surface code, stabilizers are arranged in a checkerboard pattern: $X$-type stabilizers detect $Z$ errors and $Z$-type stabilizers detect $X$ errors. Since all stabilizers commute with one another, they can be measured simultaneously within a single syndrome extraction round. The syndrome is then passed to the decoder, which infers the errors and dictates the necessary corrections. Standard surface code memory experiments typically necessitate $d$ rounds of syndrome extraction before spacetime decoding, which improves confidence in the measured syndromes \cite{Dennis_2002}.

\subsection{Surface Code With Defects}
\label{bg:defects}

Implementing the surface code on solid-state quantum processors requires large arrays of physical qubits. In practice, not all components in such arrays are operational: material defects can render individual components permanently unusable \cite{Bilmes_2020}, and transient phenomena such as two-level systems (parasitic material defects disrupting qubit coherence) \cite{Klimov_2018, M_ller_2019} or high-energy cosmic ray impacts \cite{Wu_2025} can temporarily or permanently disrupt the normal behavior of these components. We refer to inoperable qubits or couplers as \textit{defects}. 

\begin{figure}
    \centering
    \includegraphics[width=1\linewidth]{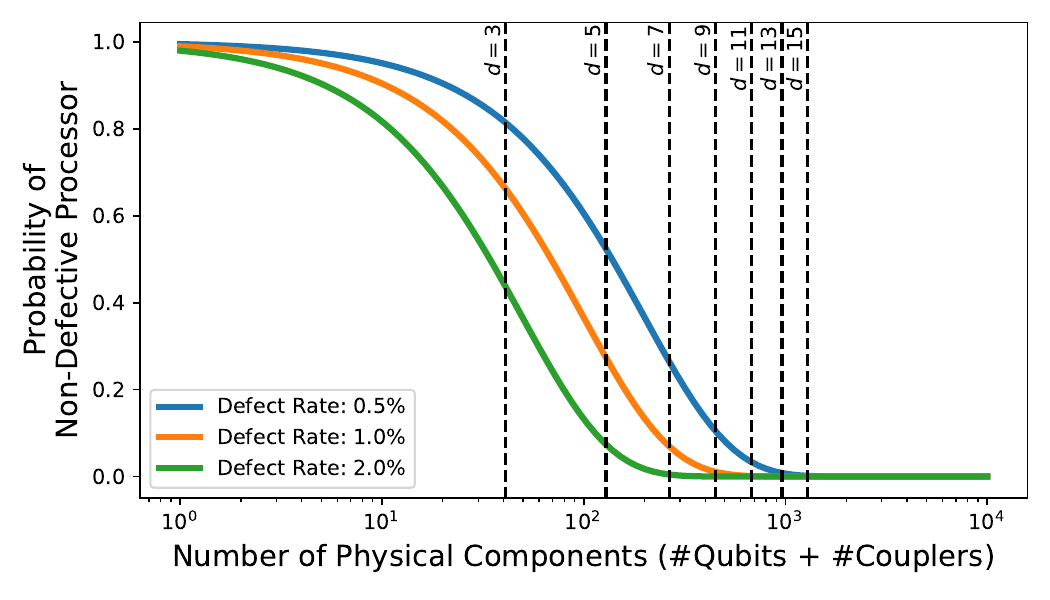}
    \caption{Probability of obtaining a non-defective processor as a function of the total number of components, for hardware defect rates of $0.5\%, 1\%$ and $2\%$. Vertical dashed lines mark the component counts corresponding to representative code distances.}
    \label{fig:defective_processor}
\end{figure}

The prevalence of such defects grows quickly with processor size. The number of physical components in a distance-$d$ surface code (including both qubits and couplers) scales as $\mathcal{O}(d^2)$. Assuming each component is independently defective with a fixed rate, Figure~\ref{fig:defective_processor} shows the probability of obtaining a non-defective processor (\textit{i.e.}, one in which all qubits and couplers are operational) as a function of the total number of components, with vertical lines marking code distances. This probability decays rapidly with system size and is effectively zero beyond a certain number of components, making large non-defective processors increasingly unlikely. 

Adapting the surface code to such defects is therefore critical for scaling quantum devices. Recent work has shown that scalable error correction remains achievable even at realistic fabrication yields \cite{PhysRevApplied.19.064081}. Early approaches addressed defects by disabling the affected components and treating them as holes in the lattice, which drastically reduced the error-correction capabilities of the code \cite{Auger_2017}. 

More recently, Snakes and Ladders (SnL)~\cite{Leroux_2025} improved on these approaches, substantially increasing the effective code distance. When a defect is encountered, SnL splits the affected stabilizer into multiple lower-weight gauge checks, which are combined to form deterministic super-stabilizers. To measure these gauge checks, SnL introduces ancilla re-purposing, in which an ancilla qubit is reassigned to measure two different gauge checks in alternating syndrome extraction rounds. Since gauge checks of opposite type ($X$ and $Z$) anti-commute, they cannot be measured within the same syndrome extraction round. This requires an alternating schedule in which $X$-type and $Z$-type super-stabilizers are measured in separate rounds, while the undamaged stabilizers are measured in every round. 

Beyond individual defect sites, SnL also introduces heuristics to generate adapted surface codes for arbitrary defect configurations. The framework is evaluated under circuit-level noise using a standard depolarizing model and the superconducting-inspired SI1000 noise model, demonstrating improvements in both effective code distance and LER relative to prior approaches \cite{Leroux_2025}.

Figure \ref{fig:surface_code_defects} shows a distance-5 surface code patch at increasing defect rates (0\% for the leftmost patch, 1\% for the middle one, and 2\% for the rightmost patch), applied to both qubits and couplers.

\subsection{Biased Noise}


Noise on quantum processors is usually asymmetric or biased. When $T_1 \gg T_2$, dephasing (Pauli $Z$) errors dominate over bit-flip (Pauli $X$) errors and vice-versa. This is characteristic of several leading quantum platforms, including trapped ion qubits \cite{PhysRevA.107.052417}, silicon spin qubits \cite{PhysRevA.109.032433, Stano_2022}, NV-center qubits \cite{PhysRevApplied.20.044045} and certain superconducting qubits \cite{PRXQuantum.4.020350, PhysRevA.106.062428}. 
For highly biased noise systems such as cat qubits, which are inherently protected against bit flips \cite{mirrahimi2014dynamically, leghtas2015confining, lescanne2020exponential}, QEC codes focusing only on phase flips are sufficient to obtain good error rates \cite{putterman2025hardware}. Our work targets the low-to-moderate bias regime, typical of many of the platforms above \cite{PhysRevA.107.052417, PhysRevA.109.032433, PhysRevA.106.062428}, in which one error type dominates but the other one cannot be ignored.

The standard surface code is suboptimal in biased settings, as it treats the $X$ and $Z$ errors symmetrically \cite{Tuckett_2018}. A line of work has therefore sought to tailor the code itself to noise asymmetry. A prominent example is the XZZX surface code, where the stabilizer checks are defined as the product of XZZX Pauli operators \cite{Bonilla_Ataides_2021}. This construction gives high thresholds for all single-qubit Pauli noise channels and demonstrates an improvement in LER scaling under biased noise. Such tailored codes have not, to our knowledge, been analyzed in the presence of defects.  

We therefore take a complementary route. As described in Section~\ref{bg:defects}, adapting the surface code to defects converts the affected stabilizers into anti-commuting gauge checks of opposite Pauli type, forcing the $X$-type and $Z$-type checks to be measured in alternating rounds. This alternating $X$/$Z$ super-stabilizers scheduling introduces an asymmetry in how the two error types are detected, exposing an important degree of freedom: the ratio of $X$ to $Z$ scheduling rounds, which serves as an additional handle without requiring any modification to the code structure. In this work, we investigate how this scheduling ratio should be chosen to match the noise bias, exploiting the defect-induced asymmetry at the scheduling level rather than re-engineering the stabilizers.

\section{Methodology}\label{sec:methodology}


\begin{figure*}
    \centering
    \includegraphics[width=\linewidth]{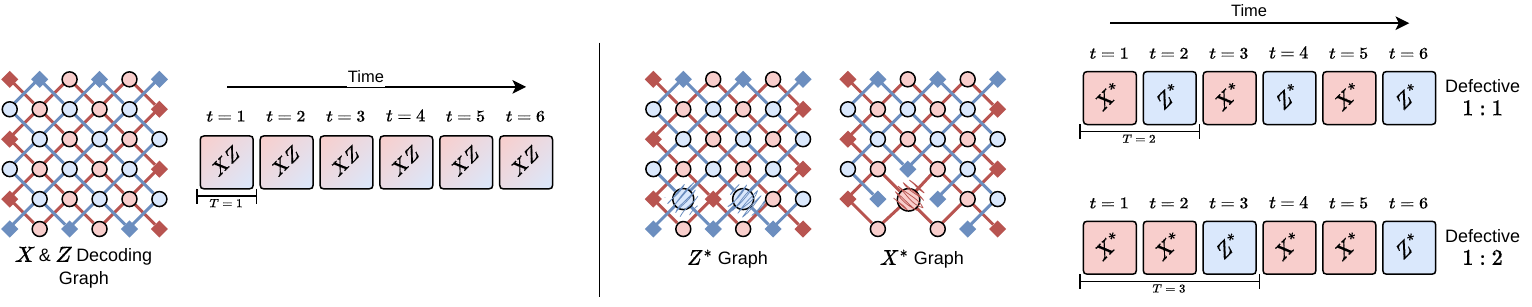}
\caption{\textbf{(left)} The $X$ \& $Z$ decoding graph for the left patch
depicted in Figure~\ref{fig:surface_code_defects}. Circular nodes represent bulk
stabilizer nodes and diamond nodes represent boundary nodes. With no defects, both the $X$ and the $Z$ stabilizers are measured in every round. \textbf{(right)} The $Z^{*}$ and $X^{*}$ decoding graphs for the two alternating syndrome extraction rounds of the middle patch in Figure~\ref{fig:surface_code_defects}, where hatched nodes mark defective stabilizers.  The non-defective $X$ and $Z$ stabilizers are measured in every round, while the $X$ and $Z$ super-stabilizers of the defective surface code are interleaved according to the scheduling ratio $R_X{:}R_Z$.}
    \label{fig:round_scheduling}
\end{figure*}

To study how the scheduling of $X$ and $Z$ syndrome extraction rounds affects the LER, we consider the rotated surface code adapted to defective qubits and couplers under biased noise. 

Defects are introduced stochastically via a single defect rate parameter $\bm{dr}$. Each qubit and each coupler on the grid is independently sampled: it is rendered defective with probability $dr$, and otherwise remains functional. For each sampled configuration, we generate the corresponding adapted surface-code patch using Snakes and Ladders (SnL) framework \cite{Leroux_2025} (Section \ref{bg:defects}). SnL guarantees that all stabilizers and super-stabilizers are measured within two error-correction rounds; with $X$- and $Z$-type gauge checks measured in alternating rounds since they anti-commute. 

We evaluate the overall LER of the system via standard surface code memory experiments, performing $\mathcal{O}(d)$ rounds of syndrome extraction for a distance $d$ surface code. We use the standard N-Z gate scheduling in each syndrome-extraction round~\cite{Fowler_2012}. All simulations are performed using Stim~\cite{gidney2021stim} under a circuit-level noise model and decoded using PyMatching~\cite{Higgott2025sparseblossom}. 

Errors are drawn from a Pauli channel with physical error rate $p$, applied after every single-qubit gate, two-qubit gate, measurement, and idle location. To study the effect of biased noise, we parametrize the asymmetry via the ratio $\bm{\eta = \frac{p_Z}{p_X}}$, where $p_Z$ and $p_X$ are the rates of Pauli $Z$ and $X$ errors, respectively. We set $p_Y=0$ throughout to isolate the per-basis effect of scheduling; $Y$ errors flip both syndromes, so including them rescales the LER by a constant factor at fixed bias, without changing the trends observed.
The symmetric (unbiased) Pauli-noise channel corresponds to $\eta=1$. To ensure a consistent noise setting across several bias ratios $\eta$, we fix the overall error added per operation to $p=p_X+p_Z$. Therefore the specific $p_X$ (and $p_Z$) for an $\eta$ noise biased system are computed as $p_X = p/(\eta+1)$ and $p_Z= p- p_X$.

As described in Section \ref{sec:background}, SnL measures $X$- and $Z$-type gauge checks in alternating rounds, with a $1{:}1$ ratio, while the undamaged stabilizers are measured every round. The uniform choice is a default, and nothing in the construction itself constrains the relative frequency of $X$- and $Z$-type rounds. 

In this work, we generalize SnL's round scheduling by defining a scheduling period $\bm{T = R_X + R_Z}$, in which $X$-type gauge checks are measured consecutively for $R_X$ rounds and $Z$-type gauge checks for $R_Z$ rounds, with undamaged stabilizers measured every round as before. We characterize a given schedule by the ratio $R = R_X{:} R_Z$; the default SnL schedule corresponds to $T = 2$ with $R = 1{:}1$. This work addresses the central question of whether departing from a uniform schedule (specifically, tuning $R$ as a function of the bias ratio $\eta$, code distance $d$, and defect rate $dr$) can improve the LER. This methodology is summarized and depicted in Figure~\ref{fig:round_scheduling}.

\section{Results}
\label{sec:results}

We now evaluate how the scheduling ratio $R$ affects the LER of defect-adapted surface codes under biased noise. Section \ref{sec:results-scheduling} establishes that non-uniform scheduling reduces the LER at a fixed bias $\eta$. Section \ref{sec:results-bias} characterizes how the optimal ratio depends on the noise bias $\eta$, and Section \ref{sec:results-defects} examines how it depends on the defect rate. 

\subsection{$X$-$Z$ Scheduling Under Biased Noise}\label{sec:results-scheduling}

\begin{figure*}
    \centering
    \includegraphics[width=1\linewidth]{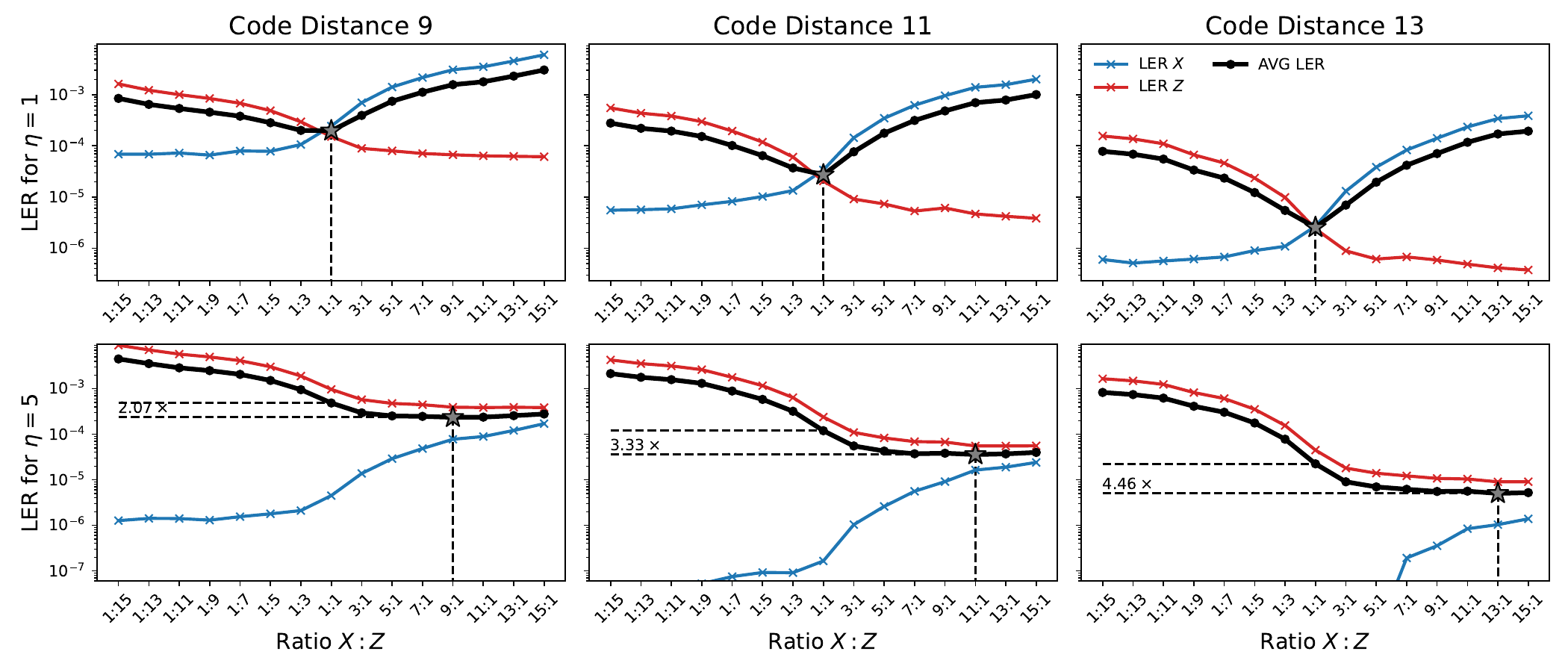}
    \caption{Logical error rate for defective systems ($dr=1\%$) for a logical qubit prepared in the $X$ basis (blue) and the $Z$ basis (red), across different scheduling ratios ($R_X{:}R_Z$), for code distances $d \in \{9, 11, 13\}$. The solid black line shows the average of the two bases. Under symmetric noise (top row), the optimum schedule is $R_X{:}R_Z = 1{:}1$; under $Z$-biased noise (bottom row), it shifts towards larger $R_X$.}
    \label{fig:z_x_LER}
\end{figure*}

We begin by characterizing how the optimal $X$-$Z$ scheduling ratio varies with noise bias in defect-adapted surface codes. Figure \ref{fig:z_x_LER} shows, for code distances $d \in \{9, 11, 13\}$, the LER for the $X$ basis (blue), the LER for the $Z$ basis (red), and the average of the two (black), evaluated over $10$ defect-adapted patches with defect rate $dr=0.01$ and physical error rate $p=10^{-3}$. 

We first validate our scheduling framework under symmetric noise ($\eta=1$). As expected, the symmetric schedule $R=1{:}1$ is optimal at every code distance (see top row of Figure \ref{fig:z_x_LER}). We next investigate biased noise at $\eta=5$, where Pauli $Z$ errors are five times more likely than Pauli $X$ errors, with the total physical error rate held fixed at $p=10^{-3}$ (see bottom row of Figure \ref{fig:z_x_LER}). 

Under biased noise, the optimal schedule shifts away from $R=1{:}1$. Allocating more rounds to $X$-type gauge measurements ($R_X > R_Z$) reduces the average LER, suppressing the $Z$ basis LER at the cost of a higher $X$ basis LER. This trade-off has a simple intuition. Under $Z$-biased noise, $Z$ errors occur more frequently, producing $X$-basis syndromes at a higher rate. If these syndromes are not detected quickly enough, uncorrected $Z$ errors accumulate and eventually flip the logical $Z$ operator. More frequent $X$-type rounds sample these syndromes sooner, directly suppressing the $Z$-basis LER. 

A second compounding effect arises from the super-stabilizers. Individual gauge-check outcomes are random, while their product (the super-stabilizer) is deterministic within a round. Repeating the same type of gauge check across consecutive rounds renders them deterministic, allowing the decoder to identify and discount measurement errors on individual gauge checks. This effect is symmetric in the $X$ and $Z$ bases: repeating either one improves the reliability of that basis's gauge measurements.

In a realistic computation, the instantaneous logical state is typically unknown, so neither basis can be preferentially protected. The relevant figure of merit is therefore the average LER over the two bases, and the optimal schedule is the one that minimizes it. Accordingly, we report only the average LER in all subsequent experiments. 

Figure \ref{fig:z_x_LER} (bottom row) shows that under biased noise ($\eta=5$), the optimal schedule improves the average LER by factors of $2.07\times$, $3.33\times$, and $4.46\times$ at code distances $d=9, 11, \text{ and } 13$, respectively. The gain from $X-Z$ scheduling grows with code distance $d$, so larger codes benefit even more from noise-adapted scheduling. 

The LER curves in the bottom row of Figure \ref{fig:z_x_LER} suggest that the average LER saturates beyond a certain $R$. We therefore define the optimal schedule as the smallest $R$ whose average LER lies within $1\%$ of the observed minimum. This tolerance accounts for variation in how defects are sampled across surface code instances and for finite-shot noise in the simulations. 

\subsection{Biased Systems and Their Impact}\label{sec:results-bias}

Biased noise arises in platforms where the relaxation time $T_1$ substantially exceeds the dephasing time $T_2$, so that phase-flip ($Z$) errors dominate over bit-flip ($X$) errors. We quantify this asymmetry as $\eta=\frac{p_Z}{p_X}$, which reduces to $\eta \approx \frac{T_1}{T_2} - \frac{1}{2}$ under the Pauli-twirling channel approximation \cite{PhysRevA.86.062318, Geller_2013}. The low-to-moderate regime $\eta \in [1, 10]$ considered in this work is characteristic of two-level superconducting and semiconductor spin-qubit platforms operated with standard dynamical-decoupling strategies, including transmon and fluxonium qubits near their coherence sweet spots \cite{Burnett_2019, PhysRevX.9.041041, PhysRevLett.130.267001}, capacitively shunted flux qubits \cite{Trappen_2025}, and silicon and germanium spin qubits \cite{Kawakami_2016, Stano_2022, RevModPhys.95.025003}. 

\begin{figure*}
    \centering
    \includegraphics[width=1\linewidth]{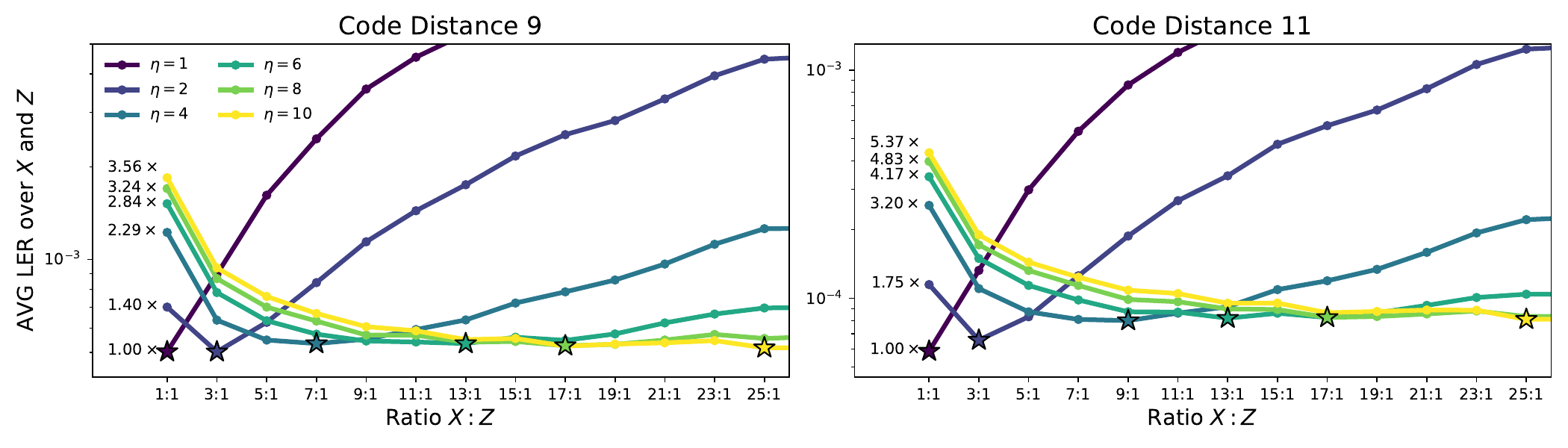}
    \caption{Average logical error rate as a function of the scheduling ratio $R$, for code distances 9 and 11 in systems with increasing bias from $\eta=1$ to $\eta=10$ ($dr=1\%$). The optimum schedule moves towards larger $R_X{:}R_Z$ as $\eta$ increases, and its improvement over $R_X{:}R_Z = 1{:}1$ grows with bias.}
    \label{fig:bias_exploration}
\end{figure*}

Understanding how the optimal scheduling ratio depends on noise bias $\eta$ would allow manufacturers and end-users to maximize processor performance by only using device calibration data \cite{mazurek2025tailoredquantumdevicecalibration, Knill_2008}, avoiding the expensive simulations otherwise required to identify the optimal ratio for each device. Figure~\ref{fig:bias_exploration} shows the average LER as a function of the scheduling ratio $R$ across bias values $\eta \in \{1, 2, 4, 6, 8, 10\}$ for code distances $d \in \{9,11\}$, evaluated over $30$ defect-adapted patches with defect rate $dr=0.01$ and physical error rate $p=10^{-3}$. As $\eta$ increases, the optimal ratio $R^*$ (depicted in the figure with a star) shifts towards larger $R_X$. The improvement in LER over $R=1{:}1$ grows with both code distance and noise bias. 

\begin{figure}
    \centering
    \includegraphics[width=1\linewidth]{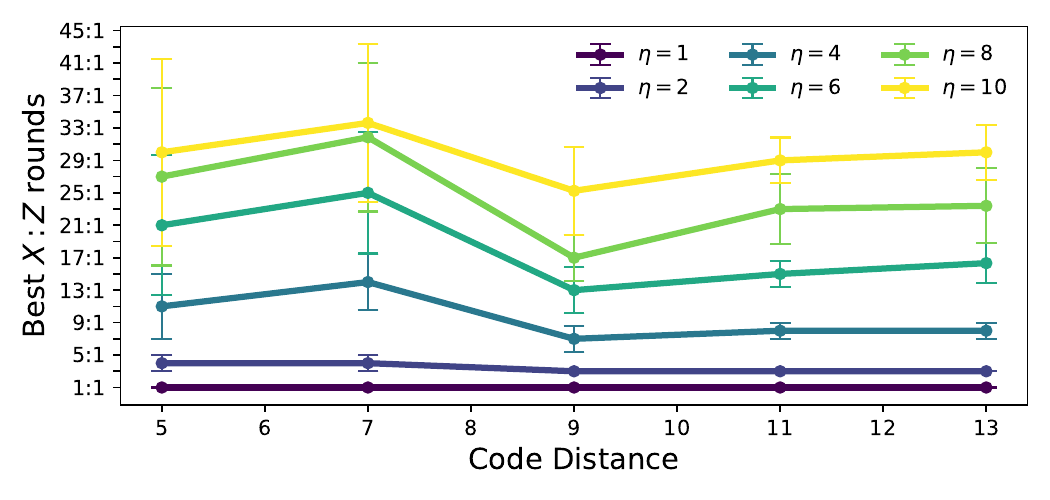}
    \caption{Optimal scheduling ratio $R^*$ as a function of code distance $d$ across noise bias $\eta \in [1, 10]$. Vertical bars indicate the range of $R$ values around $R^*$ over which the average LER stays within $5\%$ of its observed minimum; these widen with $\eta$, showing reduced sensitivity to the exact ratio at higher bias.}
    \label{fig:best_sched}
\end{figure}

To assess how the optimal schedule depends on noise bias $\eta$, Figure~\ref{fig:best_sched} shows the optimal scheduling ratio $R^*$ as a function of code distance from $d=5$ to $d=13$ across a range of bias values. The average LER fluctuates slightly around the minimum due to sampling variation. To account for this, vertical bars indicate the range of $R$ values around $R^*$ over which the average LER remains within $5\%$ of its minimum. The bars widen as $\eta$ increases, indicating that the LER is less sensitive to the exact choice of $R$ at higher bias. $R^*$ is, to a good approximation, independent of the code distance: for each $\eta$, it remains constant across $d$ up to sampling variation, and depends only on the noise bias. 

\begin{figure}
    \centering
    \includegraphics[width=1\linewidth]{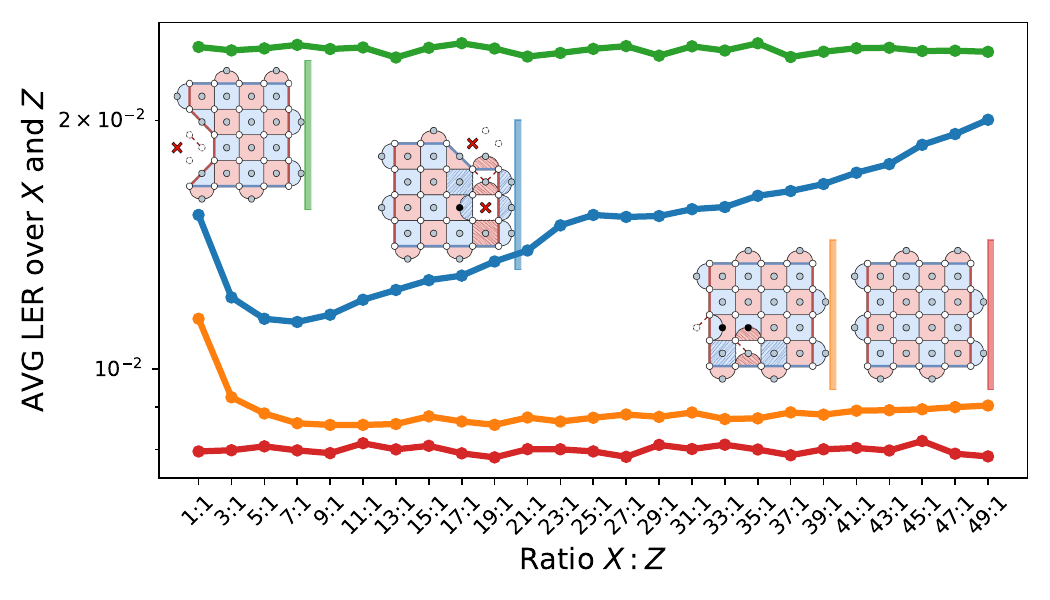}
    \caption{Average LER as a function of the scheduling ratio $R$ for four representative defect-adapted patches ($d=5$). The red sample contains no defects; the green sample contains defects, but none induce gauge operators. The blue and orange samples both contain gauge-inducing defects, with the orange sample preserving the code distance and the blue one reducing it. Samples without gauge checks are excluded from the scheduling analysis, since round scheduling is meaningful only when gauge checks are present.}
    \label{fig:defects_variability}
\end{figure}

At smaller distances ($d \in \{5, 7\}$), some surface code samples contain no defects. At a 1\% defect rate, the small number of qubits and couplers in these samples occasionally yields a defect-free processor. Other samples may consist solely of intact stabilizers without gauge checks, rendering round scheduling unnecessary. Even when gauge checks are present, the location and type of defect (qubit or coupler) reduce the effective code distance by varying amounts, further affecting the LER. Figure \ref{fig:defects_variability} illustrates these cases for the $d=5$ code. Since scheduling is meaningful only when gauge checks are present, we exclude samples with no defects or no gauge checks from our analysis. 

Intuitively, $R^*$ balances the rates at which $X$- and $Z$-type gauge checks are measured, and this balance is set by the ratio of physical error rates, not by the number of physical qubits encoding the logical qubit. Scaling the code changes the LER at a fixed schedule, but not the location of the minimum. A manufacturer can therefore determine $R^*$ once from device calibration data and reuse it across all code sizes. 


\subsection{Impact of Defect Rate}\label{sec:results-defects}

Lastly, we study how the optimal scheduling ratio $R^*$ varies as a function of the defect rate $dr$. A higher defect rate produces more super-stabilizers on the adapted surface-code patch, increasing the number of gauge checks subject to the alternating schedule.

\begin{figure*}
    \centering
    \includegraphics[width=1\linewidth]{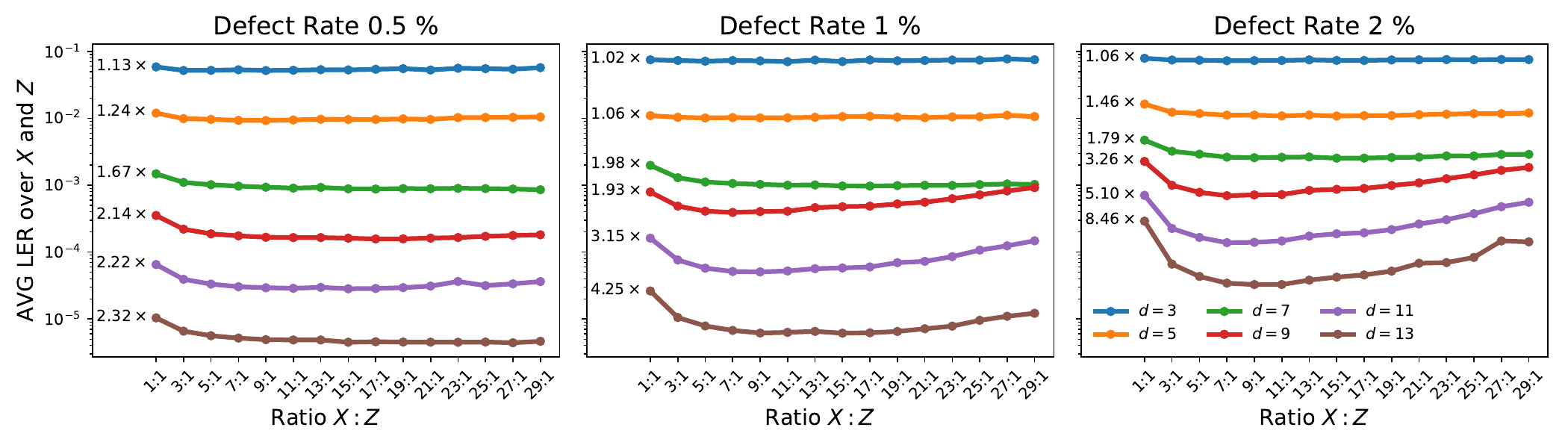}
    \caption{Average LER as a function of the scheduling ratio R for code distances from $d=3$ to $d=13$ and architectures with increasing defect rate, at fixed noise bias $\eta=5$. The logical error rate rises with $dr$ at every distance and ratio.}
    \label{fig:vary_dr}
\end{figure*}

As expected, the improvement in LER of the optimal schedule over the $R=1{:}1$ baseline grows with defect rate $dr$, reaching $8.46\times$ at code distance $d=13$ for systems with a defect rate of $2\%$. The average LER itself also rises with $dr$ across every code distance and scheduling ratio we tested. This means that, despite the gains from $X$-$Z$ checks scheduling, minimizing defects at the manufacturing stage remains critical for achieving low LERs. 

Figure \ref{fig:vary_dr} shows that the optimal schedule is essentially independent of $dr$. Although larger $dr$ gives the alternating schedule more super-stabilizers to act on, $R^*$ remains effectively unchanged across the range tested. Combined with the earlier finding (Figure \ref{fig:best_sched}) that $R^*$ is set primarily by the noise bias and is independent of $d$, this shows that once $\eta$ is obtained from device calibration data, the resulting $R^*$ serves every code size. 


\section{Application of $X$-$Z$ Scheduling to Non-Defective Architectures}
\label{sec:crosstalk}

\begin{figure}
    \centering
    \includegraphics[width=1\linewidth]{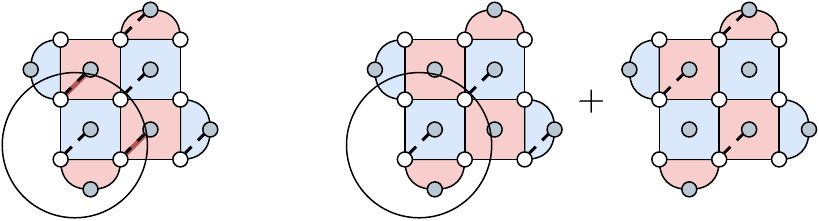}
    \caption{\textbf{(left)} When $X$- and $Z$-type stabilizers are measured in the same round, each layer of the syndrome-extraction circuit executes one CNOT per stabilizer in parallel, leading to crosstalk errors between them. \textbf{(right)} Scheduling a single stabilizer type per round halves the number of parallel CNOTs per layer, suppressing the per-CNOT crosstalk error.} 
    \label{fig:crosstalk_method}
\end{figure}

The round-scheduling degree of freedom in defect-adapted codes arises from anti-commuting gauge checks. In a defect-free surface code, the $X$- and $Z$-type stabilizers commute and can be measured in the same syndrome extraction round, so there is no need for round scheduling. In this section, however, we show that the \textit{benefits of separating $X$- and $Z$-type stabilizer rounds extend to defect-free architectures whenever crosstalk between simultaneously executed CNOT gates is a dominant error source}. 

In quantum computers, crosstalk noise refers to a broad class of phenomena in which one subsystem unintentionally affects another \cite{Sarovar_2020}. In transmon-based systems, this commonly takes the form of residual couplings between transmons that should be uncoupled. More specifically, crosstalk may arise when a control signal targeting one qubit affects adjacent qubits, when qubits operating at the same frequency interact unintentionally \cite{11250000}, or when parallel instructions induce noise on one another \cite{Murali_2020}. 

No standardized model of crosstalk noise exists across hardware platforms. Drawing on prior work and benchmarking data \cite{Murali_2020, 11250000, Wagner_2025}, we adopt a simultaneous-execution model in which pairs of CNOT gates induce additional noise on one another when executed in parallel. This model is a simple abstraction of an experimentally observed effect of crosstalk \cite{Murali_2020}.

Figure \ref{fig:crosstalk_method} (left) illustrates this for a $d=3$ surface-code syndrome extraction round: all CNOTs in a layer are executed in parallel and therefore suffer from crosstalk noise. Separating $X$- and $Z$-type stabilizer measurements into distinct rounds halves the number of parallel two-qubit gates per layer and suppresses the per-CNOT crosstalk error (Figure \ref{fig:crosstalk_method} (right)), at the cost of leaving the inactive qubits idle for the duration of the round. This raises the question: can round scheduling reduce the LER of a defect-free surface code operating under biased noise and crosstalk? 

To quantify the crosstalk strength, we follow prior simultaneous randomized benchmarking studies \cite{Wagner_2025, niu2021analyzingcrosstalkerrornisq}, which report the ratio of simultaneous gate error rate to isolated gate error rate for two gates executed in parallel. Denoting this ratio by $\alpha$, we vary it in the range $1 \leq \alpha \leq 2$. CNOTs are assigned a baseline error rate of $p=10^{-3}$ when executed in isolation and an inflated rate of $\alpha \cdot p$ when executed simultaneously. Here, $\alpha=1$ corresponds to the crosstalk-free case. Single-qubit gates, measurements, and qubit idling are modeled with biased noise at rate $p=10^{-3}$ in all cases. All qubits accumulate idling noise during every timestep in which they are not involved in a gate. Throughout all experiments, we fix the noise bias $\eta=5$. 

Under this model, any syndrome extraction round that measures $X$- and $Z$-type stabilizers together incurs the crosstalk-inflated CNOT rate $\alpha \cdot p$, whereas rounds measuring only $X$-type (or only $Z$-type) stabilizers incur the baseline error rate $10^{-3}$. Non-uniform scheduling should therefore help on two fronts under $Z$-biased noise: $X$-type rounds sample the dominant errors more frequently and avoid the crosstalk penalty of parallel execution. The following experiments quantify this combined effect.

We run surface-code memory experiments over $26$ syndrome-measurement rounds at noise bias $\eta=5$, sweeping code distances $d\in \{3,5,7,9,11,13\}$, and crosstalk ratios $\alpha \in [1, 2]$. We consider two parametrized schedules: $R=1{:}R_X$, where each scheduling period of length $T=1+R_X$ consists of one round measuring all stabilizers ($X$ and $Z$ with a CNOT error rate of $\alpha \cdot p$) followed by $R_X$ rounds measuring only $X$-type stabilizers; and its counterpart $R=1{:}R_Z$, with $Z$-type rounds instead. Each period is repeated for the full $26$ rounds.

\begin{figure*}
    \centering
    \includegraphics[width=1\linewidth]{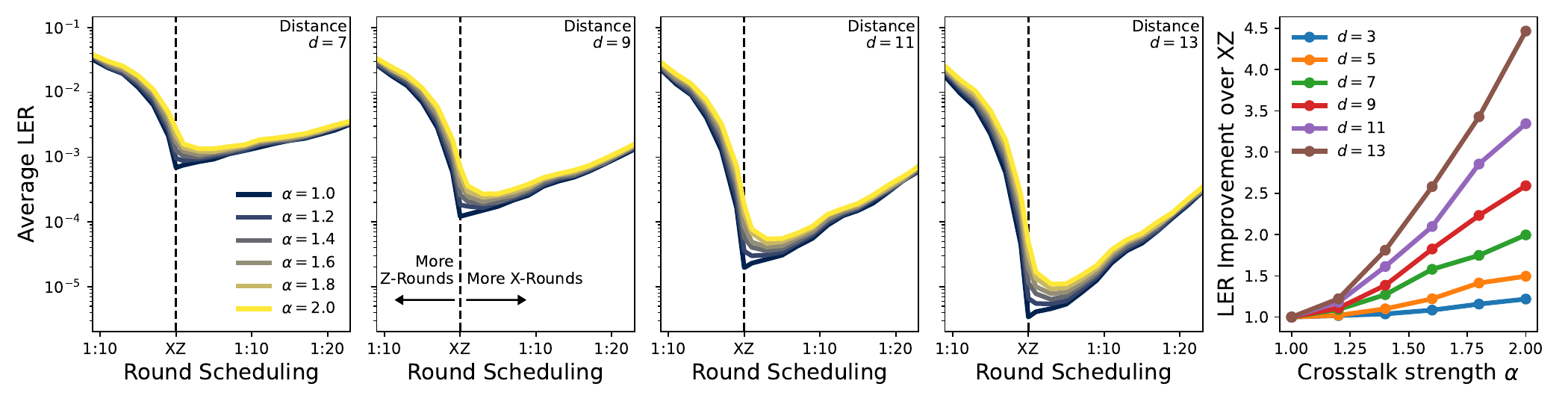}
    \caption{\textbf{(left)} Average LER as a function of round scheduling. The central tick $XZ$ denotes the baseline in which all stabilizers are measured in every round; points to its left correspond to $1{:}R_X$ schedules and points to its right correspond to $1{:}R_Z$ schedules. \textbf{(right)} LER improvement relative to the baseline for $\alpha \in [1,2]$, showing that the gain grows with both the code distance $d$ and the crosstalk ratio $\alpha$.}
    \label{fig:crosstalk_results}
\end{figure*}

Under $Z$-biased noise, only the $1{:}R_X$ schedule is beneficial: extra $X$-type rounds both sample the dominant $Z$ errors more frequently and avoid the crosstalk penalty of parallel execution. Figure \ref{fig:crosstalk_results} shows that the optimal $1{:}R_X$ schedule departs from the baseline $R_X=0$ (all mixed rounds) for every $(d, \alpha)$ configuration with $\alpha > 1$, and the LER improvement grows with both the code distance $d$ and the crosstalk ratio $\alpha$. Average LER improves up to $4.5\times$ for a system with code distance $13$, a crosstalk factor of $\alpha=2$, and noise bias of $\eta=5$.

\section{Related work}\label{sec:related work}

Two lines of work motivate this paper: frameworks for adapting the surface code to defects and codes designed for asymmetric Pauli-noise channels. The two have developed independently: defect-adapted frameworks are benchmarked against symmetric noise channels, and bias-tailored codes are analyzed on defect-free lattices. 

\subsection{Adapting the Surface Code to Defects}

Several frameworks have been proposed for adapting the surface code to defective qubits and couplers, all evaluated under standard depolarizing or SI1000 noise models. The earliest proposals \cite{Auger_2017, Siegel_2023} define what we refer to as Data Qubit Disabling (DQD) strategy, following the terminology of Leroux et al.\cite{Leroux_2025}. DQD disables any defective data qubit, together with every data qubit surrounding a defective ancilla or coupler. It then converts the stabilizers around these disabled data qubits into gauge checks, whose products form super-stabilizers. This approach unavoidably reduces the code distance by two for a defective ancilla qubit and by one for a defective coupler, resulting in a higher LER that increases with the number of defects in the system.

Snakes and Ladders (SnL) \cite{Leroux_2025} instead reuses neighboring ancillas to form gauge checks, typically preserving the code distance where DQD loses it. We describe SnL in detail in Section~\ref{sec:background}, as it is the framework on which we build.

LUCI \cite{Debroy_2025} takes a different approach, rooted in the mid-cycle state of the surface code -- an unrotated surface code on both data and measure qubits \cite{McEwen_2023}. The LUCI framework constructs fault-tolerant circuits from a small set of rounds, each starting and ending at the mid-cycle state, with individual stabilizer measurements described by L-, U-, C-, and I- shaped sub-circuits. Applied to defect adaptation, LUCI preserves the spacelike distance at the cost of halving the timelike distance and measures all stabilizers within four rounds. 

Automated compilation including dropouts (ACID) \cite{wolanski2026automatedcompilationincludingdropouts} is a more recent framework that operates in an ancilla-free regime, where no qubits are pre-assigned as measurement ancillas, and stabilizers can be measured using any supporting data qubits. ACID generalizes LUCI and compiles syndrome-measurement circuits for any CSS stabilizer code under arbitrary dropout patterns and device connectivity.

Minimally invasive alterations (MIA) \cite{mishmash2025excisingdeadcomponentssurface} provides numerical simulations of a defect adaptation strategy built using the pairwise measurement-based surface code \cite{Grans_Samuelsson_2024}, which is highly optimized for Majorana-based hardware. MIA decomposes stabilizer checks into two-qubit parity measurements and pipelines $X$- and $Z$-type gauge checks within the same syndrome extraction round. The decoder is also modified to handle pairwise-measurement circuits. MIA is compared against a pairwise-measurement implementation of DQD \cite{Auger_2017} and has not yet been implemented in a CNOT-based setting. 

\subsection{QEC Under Biased Noise}

A parallel line of work designs codes that exploit noise bias to achieve higher thresholds and improved LERs on real hardware. We restrict attention to surface-code variants. 

Bias-tailored surface codes \cite{Tuckett_2018} apply a simple modification to the stabilizer structure of the standard surface code, where they exchange the $Z$-type stabilizers with products of $Y$ around each plaquette. This yields a very high threshold of $43.7\%$ using a tensor-network-based decoder under pure dephasing noise.

The XZZX surface code \cite{Bonilla_Ataides_2021} is another variant of the standard surface code where the stabilizer checks are given by the product of XZZX Pauli operators around each face on the square lattice. This code performs well across a wide range of bias values and has been widely adopted. 

Clifford deformed surface codes (CDSCs) \cite{Dua_2024} are obtained by applying single-qubit Clifford operators to each qubit of a standard surface code. Dua et al. show that random Clifford deformations can outperform the XZZX surface codes at finite noise bias values, and that further gains from CDSCs remain possible through more targeted deformation choices.

\section{Conclusions and Future Work}\label{sec:conclusion}

We have shown that the alternating $X$/$Z$ check schedule imposed by defect adaptation using SnL \cite{Leroux_2025} is an exploitable design parameter. Under biased noise, the uniform schedule $R=1{:}1$ is suboptimal: reallocating rounds towards the basis that detects the dominant error type reduces the average LER by up to $4.25\times$ at $\eta=5$ and $d=13$ under a $1\%$ defect rate, and up to $8.46\times$ at $d=13$ under a $2\%$ defect rate. The optimal ratio is, to a good approximation, independent of the code distance and is set primarily by the noise bias $\eta$. This allows manufacturers to select the scheduling ratio directly from device calibration data, without running code-specific simulations. 

\begin{figure}
    \centering
    \includegraphics[width=1\linewidth]{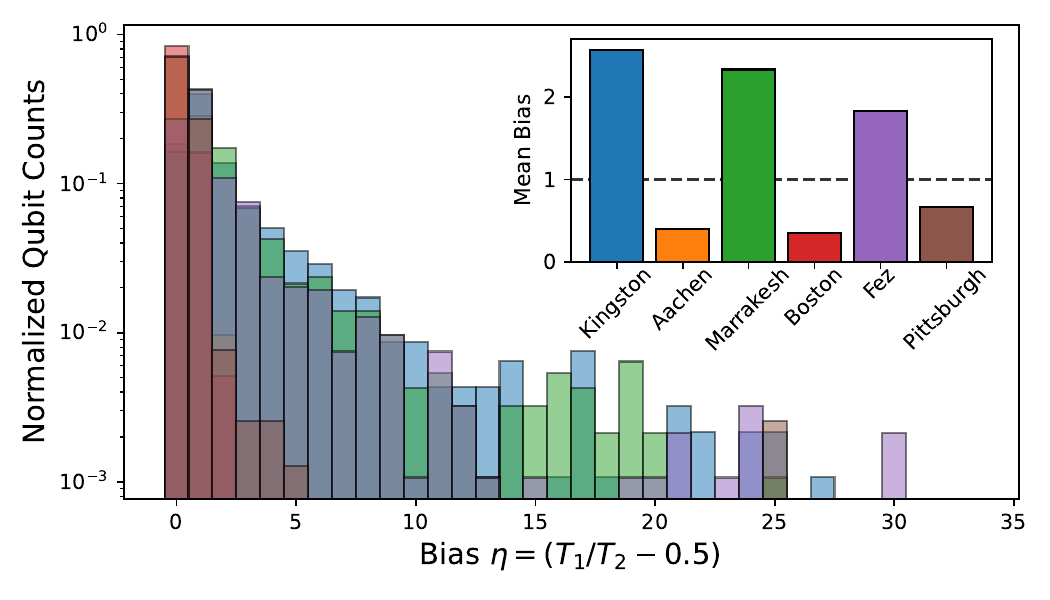}
    \caption{Noise bias $\eta$ on publicly available IBM Heron processors, extracted from reported $T_1$ and $T_2$ calibration data, via $\eta \approx T_1/T_2 - 1/2$. The main panel shows the distribution of $\eta$ over qubits for each device, and the inset reports the per-device mean.}
    \label{fig:calibration}
\end{figure}

Several directions remain open. First, bias-tailored codes such as the XZZX code and Clifford-deformed surface codes redesign the stabilizers; how these constructions can be adapted to defects is an open question. A systematic comparison between bias adaptation at the stabilizer level and at the scheduling level (the approach of this paper) would help us better understand when each is preferable, and whether the two can be composed. Second, we have focused on the surface code, but the alternating-schedule degree of freedom is not unique to it. Any CSS code with a defect-adaptation procedure that produces anti-commuting gauge checks would have a similar scheduling parameter. Third, real devices can also exhibit substantial qubit-to-qubit variation in noise bias. For instance, we compute the noise bias $\eta \approx \frac{T_1}{T_2}-\frac{1}{2}$ from publicly available calibration data of IBM's Heron processors \cite{IBMQuantum} in Figure \ref{fig:calibration}. The histogram aggregates data over $5$ days for $156$ qubits on each processor.  Three of the processors exhibit $X$-biased noise ($\eta < 1$), while the remaining $Z$-biased processors have per-device mean biases up to $\eta=2.6$. An interesting idea to explore would be to allow the scheduling ratio to match the local bias of a region, which would require modifications to multiple parts of the pipeline but could deliver further gains on heterogeneous processors. Finally, beyond the trends shown in this work, a device-specific noise model that accounts for noise bias, coherence times, and crosstalk would enable an in-depth characterization of the scheduling parameters that maximize the performance of a specific processor.

More broadly, defect adaptation has an associated cost -- reduced distance and additional syndrome extraction rounds due to gauge checks. Our results suggest that the alternating gauge-checks schedule it requires is itself a design parameter, and under biased noise, it becomes a useful optimization parameter to improve the LER. The only input required to find this scheduling is device calibration data. As quantum processors scale and defects become unavoidable, turning the defect-induced constraints into tunable parameters will be increasingly valuable.

\section*{Acknowledgment}

This article is based in part upon work supported by the U.S. Department of Energy, Office of Science, National Quantum Information Science Research Centers, and Co-design Center for Quantum Advantage (C2QA) under contract DESC0012704.


\bibliographystyle{IEEEtran}
\bibliography{references}

@article{Debroy_2025,
   title={LUCI in the Surface Code with Dropouts},
   volume={9},
   ISSN={2521-327X},
   journal={Quantum},
   publisher={Verein zur Forderung des Open Access Publizierens in den Quantenwissenschaften},
   author={Debroy, Dripto M. and McEwen, Matt and Gidney, Craig and Shutty, Noah and Zalcman, Adam},
   year={2025},
   month=dec, pages={1936} }

@article{Leroux_2025,
   title={Snakes and Ladders: Adapting the Surface Code to Defects},
   volume={6},
   ISSN={2691-3399},
   number={4},
   journal={PRX Quantum},
   publisher={American Physical Society (APS)},
   author={Leroux, Catherine and Lin, Sophia F. and Bienias, Przemyslaw and Sankar, Krishanu R. and Benhemou, Asmae and Kubica, Aleksander and Iverson, Joseph K.},
   year={2025},
   month=oct }

@article{gidney2021stim,
  title = {Stim: a fast stabilizer circuit simulator},
  author = {Gidney, Craig},
  journal = {{Quantum}},
  issn = {2521-327X},
  publisher = {{Verein zur F{\"{o}}rderung des Open Access Publizierens
                in den Quantenwissenschaften}},
  volume = 5,
  pages = 497,
  month = jul,
  year = 2021
}

@article{Fowler_2012,
   title={Surface codes: Towards practical large-scale quantum computation},
   volume={86},
   ISSN={1094-1622},
   number={3},
   journal={Physical Review A},
   publisher={American Physical Society (APS)},
   author={Fowler, Austin G. and Mariantoni, Matteo and Martinis, John M. and Cleland, Andrew N.},
   year={2012},
   month=sep }

@article{Auger_2017,
   title={Fault-tolerance thresholds for the surface code with fabrication errors},
   volume={96},
   ISSN={2469-9934},
   number={4},
   journal={Physical Review A},
   publisher={American Physical Society (APS)},
   author={Auger, James M. and Anwar, Hussain and Gimeno-Segovia, Mercedes and Stace, Thomas M. and Browne, Dan E.},
   year={2017},
   month=oct }

@article{PhysRevApplied.19.064081,
  title = {Quantum Computing is Scalable on a Planar Array of Qubits with Fabrication Defects},
  author = {Strikis, Armands and Benjamin, Simon C. and Brown, Benjamin J.},
  journal = {Phys. Rev. Appl.},
  volume = {19},
  issue = {6},
  pages = {064081},
  numpages = {18},
  year = {2023},
  month = {Jun},
  publisher = {American Physical Society},
}

@article{Tuckett_2018,
   title={Ultrahigh Error Threshold for Surface Codes with Biased Noise},
   volume={120},
   ISSN={1079-7114},
   number={5},
   journal={Physical Review Letters},
   publisher={American Physical Society (APS)},
   author={Tuckett, David K. and Bartlett, Stephen D. and Flammia, Steven T.},
   year={2018},
   month=jan }

@article{Bonilla_Ataides_2021,
   title={The XZZX surface code},
   volume={12},
   ISSN={2041-1723},
   number={1},
   journal={Nature Communications},
   publisher={Springer Science and Business Media LLC},
   author={Bonilla Ataides, J. Pablo and Tuckett, David K. and Bartlett, Stephen D. and Flammia, Steven T. and Brown, Benjamin J.},
   year={2021},
   month=apr }

@article{Siegel_2023,
   title={Adaptive surface code for quantum error correction in the presence of temporary or permanent defects},
   volume={7},
   ISSN={2521-327X},
   journal={Quantum},
   publisher={Verein zur Forderung des Open Access Publizierens in den Quantenwissenschaften},
   author={Siegel, Adam and Strikis, Armands and Flatters, Thomas and Benjamin, Simon},
   year={2023},
   month=jul, pages={1065} }

@article{Dua_2024,
   title={Clifford-Deformed Surface Codes},
   volume={5},
   ISSN={2691-3399},
   number={1},
   journal={PRX Quantum},
   publisher={American Physical Society (APS)},
   author={Dua, Arpit and Kubica, Aleksander and Jiang, Liang and Flammia, Steven T. and Gullans, Michael J.},
   year={2024},
   month=mar }

@article{McEwen_2023,
   title={Relaxing Hardware Requirements for Surface Code Circuits using Time-dynamics},
   volume={7},
   ISSN={2521-327X},
   journal={Quantum},
   publisher={Verein zur Forderung des Open Access Publizierens in den Quantenwissenschaften},
   author={McEwen, Matt and Bacon, Dave and Gidney, Craig},
   year={2023},
   month=nov, pages={1172} }

@article{wolanski2026automatedcompilationincludingdropouts,
  title={Automated Compilation Including Dropouts: Tolerating Defective Components in Stabiliser Codes},
  author={Wolanski, Stasiu},
  journal={arXiv preprint arXiv:2512.01943},
  year={2025}
}

@article{mishmash2025excisingdeadcomponentssurface,
  title={Excising dead components in the surface code using minimally invasive alterations: A performance study},
  author={Mishmash, Ryan V and Kliuchnikov, Vadym and Bello-Rivas, Juan and Paetznick, Adam and Aasen, David and Knapp, Christina and Wu, Yue and Bauer, Bela and da Silva, Marcus P and Bonderson, Parsa},
  journal={arXiv preprint arXiv:2508.04786},
  year={2025}
}

@article{Grans_Samuelsson_2024,
   title={Improved Pairwise Measurement-Based Surface Code},
   volume={8},
   ISSN={2521-327X},
   journal={Quantum},
   publisher={Verein zur Forderung des Open Access Publizierens in den Quantenwissenschaften},
   author={Grans-Samuelsson, Linnea and Mishmash, Ryan V. and Aasen, David and Knapp, Christina and Bauer, Bela and Lackey, Brad and Silva, Marcus P. da and Bonderson, Parsa},
   year={2024},
   month=aug, pages={1429} }

@article{Higgott2025sparseblossom,
  title = {Sparse {B}lossom: correcting a million errors per core second with minimum-weight matching},
  author = {Higgott, Oscar and Gidney, Craig},
  journal = {{Quantum}},
  issn = {2521-327X},
  publisher = {{Verein zur F{\"{o}}rderung des Open Access Publizierens in den Quantenwissenschaften}},
  volume = {9},
  pages = {1600},
  month = jan,
  year = {2025}
}

@article{Bilmes_2020,
   title={Resolving the positions of defects in superconducting quantum bits},
   volume={10},
   ISSN={2045-2322},
   number={1},
   journal={Scientific Reports},
   publisher={Springer Science and Business Media LLC},
   author={Bilmes, Alexander and Megrant, Anthony and Klimov, Paul and Weiss, Georg and Martinis, John M. and Ustinov, Alexey V. and Lisenfeld, Jürgen},
   year={2020},
   month=feb }

@article{Klimov_2018,
   title={Fluctuations of Energy-Relaxation Times in Superconducting Qubits},
   volume={121},
   ISSN={1079-7114},
   number={9},
   journal={Physical Review Letters},
   publisher={American Physical Society (APS)},
   author={Klimov, P. V. and Kelly, J. and Chen, Z. and Neeley, M. and Megrant, A. and Burkett, B. and Barends, R. and Arya, K. and Chiaro, B. and Chen, Yu and Dunsworth, A. and Fowler, A. and Foxen, B. and Gidney, C. and Giustina, M. and Graff, R. and Huang, T. and Jeffrey, E. and Lucero, Erik and Mutus, J. Y. and Naaman, O. and Neill, C. and Quintana, C. and Roushan, P. and Sank, Daniel and Vainsencher, A. and Wenner, J. and White, T. C. and Boixo, S. and Babbush, R. and Smelyanskiy, V. N. and Neven, H. and Martinis, John M.},
   year={2018},
   month=aug }

@article{M_ller_2019,
   title={Towards understanding two-level-systems in amorphous solids: insights from quantum circuits},
   volume={82},
   ISSN={1361-6633},
   number={12},
   journal={Reports on Progress in Physics},
   publisher={IOP Publishing},
   author={Müller, Clemens and Cole, Jared H and Lisenfeld, Jürgen},
   year={2019},
   month=oct, pages={124501} }

@article{Aliferis_2008,
   title={Fault-tolerant quantum computation against biased noise},
   volume={78},
   ISSN={1094-1622},
   number={5},
   journal={Physical Review A},
   publisher={American Physical Society (APS)},
   author={Aliferis, Panos and Preskill, John},
   year={2008},
   month=nov }

@article{putterman2025hardware,
  title={Hardware-efficient quantum error correction via concatenated bosonic qubits},
  author={Putterman, Harald and Noh, Kyungjoo and Hann, Connor T and MacCabe, Gregory S and Aghaeimeibodi, Shahriar and Patel, Rishi N and Lee, Menyoung and Jones, William M and Moradinejad, Hesam and Rodriguez, Roberto and others},
  journal={Nature},
  volume={638},
  number={8052},
  pages={927--934},
  year={2025},
  publisher={Nature Publishing Group UK London}
}

@article{mirrahimi2014dynamically,
  title={Dynamically protected cat-qubits: a new paradigm for universal quantum computation},
  author={Mirrahimi, Mazyar and Leghtas, Zaki and Albert, Victor V and Touzard, Steven and Schoelkopf, Robert J and Jiang, Liang and Devoret, Michel H},
  journal={New Journal of Physics},
  volume={16},
  number={4},
  pages={045014},
  year={2014},
  publisher={IOP Publishing}
}

@article{leghtas2015confining,
  title={Confining the state of light to a quantum manifold by engineered two-photon loss},
  author={Leghtas, Zaki and Touzard, Steven and Pop, Ioan M and Kou, Angela and Vlastakis, Brian and Petrenko, Andrei and Sliwa, Katrina M and Narla, Anirudh and Shankar, Shyam and Hatridge, Michael J and others},
  journal={Science},
  volume={347},
  number={6224},
  pages={853--857},
  year={2015},
  publisher={American Association for the Advancement of Science}
}

@article{lescanne2020exponential,
  title={Exponential suppression of bit-flips in a qubit encoded in an oscillator},
  author={Lescanne, Rapha{\"e}l and Villiers, Marius and Peronnin, Th{\'e}au and Sarlette, Alain and Delbecq, Matthieu and Huard, Benjamin and Kontos, Takis and Mirrahimi, Mazyar and Leghtas, Zaki},
  journal={Nature Physics},
  volume={16},
  number={5},
  pages={509--513},
  year={2020},
  publisher={Nature Publishing Group UK London}
}

@article{2024,
   title={Quantum error correction below the surface code threshold},
   volume={638},
   ISSN={1476-4687},
   number={8052},
   journal={Nature},
   publisher={Springer Science and Business Media LLC},
   author={Acharya, Rajeev and Abanin, Dmitry A. and Aghababaie-Beni, Laleh and Aleiner, Igor and Andersen, Trond I. and Ansmann, Markus and Arute, Frank and Arya, Kunal and Asfaw, Abraham and Astrakhantsev, Nikita and Atalaya, Juan and Babbush, Ryan and Bacon, Dave and Ballard, Brian and Bardin, Joseph C. and Bausch, Johannes and Bengtsson, Andreas and Bilmes, Alexander and Blackwell, Sam and Boixo, Sergio and Bortoli, Gina and Bourassa, Alexandre and Bovaird, Jenna and Brill, Leon and Broughton, Michael and Browne, David A. and Buchea, Brett and Buckley, Bob B. and Buell, David A. and Burger, Tim and Burkett, Brian and Bushnell, Nicholas and Cabrera, Anthony and Campero, Juan and Chang, Hung-Shen and Chen, Yu and Chen, Zijun and Chiaro, Ben and Chik, Desmond and Chou, Charina and Claes, Jahan and Cleland, Agnetta Y. and Cogan, Josh and Collins, Roberto and Conner, Paul and Courtney, William and Crook, Alexander L. and Curtin, Ben and Das, Sayan and Davies, Alex and De Lorenzo, Laura and Debroy, Dripto M. and Demura, Sean and Devoret, Michel and Di Paolo, Agustin and Donohoe, Paul and Drozdov, Ilya and Dunsworth, Andrew and Earle, Clint and Edlich, Thomas and Eickbusch, Alec and Elbag, Aviv Moshe and Elzouka, Mahmoud and Erickson, Catherine and Faoro, Lara and Farhi, Edward and Ferreira, Vinicius S. and Burgos, Leslie Flores and Forati, Ebrahim and Fowler, Austin G. and Foxen, Brooks and Ganjam, Suhas and Garcia, Gonzalo and Gasca, Robert and Genois, {\'E}lie and Giang, William and Gidney, Craig and Gilboa, Dar and Gosula, Raja and Dau, Alejandro Grajales and Graumann, Dietrich and Greene, Alex and Gross, Jonathan A. and Habegger, Steve and Hall, John and Hamilton, Michael C. and Hansen, Monica and Harrigan, Matthew P. and Harrington, Sean D. and Heras, Francisco J. H. and Heslin, Stephen and Heu, Paula and Higgott, Oscar and Hill, Gordon and Hilton, Jeremy and Holland, George and Hong, Sabrina and Huang, Hsin-Yuan and Huff, Ashley and Huggins, William J. and Ioffe, Lev B. and Isakov, Sergei V. and Iveland, Justin and Jeffrey, Evan and Jiang, Zhang and Jones, Cody and Jordan, Stephen and Joshi, Chaitali and Juhas, Pavol and Kafri, Dvir and Kang, Hui and Karamlou, Amir H. and Kechedzhi, Kostyantyn and Kelly, Julian and Khaire, Trupti and Khattar, Tanuj and Khezri, Mostafa and Kim, Seon and Klimov, Paul V. and Klots, Andrey R. and Kobrin, Bryce and Kohli, Pushmeet and Korotkov, Alexander N. and Kostritsa, Fedor and Kothari, Robin and Kozlovskii, Borislav and Kreikebaum, John Mark and Kurilovich, Vladislav D. and Lacroix, Nathan and Landhuis, David and Lange-Dei, Tiano and Langley, Brandon W. and Laptev, Pavel and Lau, Kim-Ming and Le Guevel, Loïck and Ledford, Justin and Lee, Joonho and Lee, Kenny and Lensky, Yuri D. and Leon, Shannon and Lester, Brian J. and Li, Wing Yan and Li, Yin and Lill, Alexander T. and Liu, Wayne and Livingston, William P. and Locharla, Aditya and Lucero, Erik and Lundahl, Daniel and Lunt, Aaron and Madhuk, Sid and Malone, Fionn D. and Maloney, Ashley and Mandrà, Salvatore and Manyika, James and Martin, Leigh S. and Martin, Orion and Martin, Steven and Maxfield, Cameron and McClean, Jarrod R. and McEwen, Matt and Meeks, Seneca and Megrant, Anthony and Mi, Xiao and Miao, Kevin C. and Mieszala, Amanda and Molavi, Reza and Molina, Sebastian and Montazeri, Shirin and Morvan, Alexis and Movassagh, Ramis and Mruczkiewicz, Wojciech and Naaman, Ofer and Neeley, Matthew and Neill, Charles and Nersisyan, Ani and Neven, Hartmut and Newman, Michael and Ng, Jiun How and Nguyen, Anthony and Nguyen, Murray and Ni, Chia-Hung and Niu, Murphy Yuezhen and O’Brien, Thomas E. and Oliver, William D. and Opremcak, Alex and Ottosson, Kristoffer and Petukhov, Andre and Pizzuto, Alex and Platt, John and Potter, Rebecca and Pritchard, Orion and Pryadko, Leonid P. and Quintana, Chris and Ramachandran, Ganesh and Reagor, Matthew J. and Redding, John and Rhodes, David M. and Roberts, Gabrielle and Rosenberg, Eliott and Rosenfeld, Emma and Roushan, Pedram and Rubin, Nicholas C. and Saei, Negar and Sank, Daniel and Sankaragomathi, Kannan and Satzinger, Kevin J. and Schurkus, Henry F. and Schuster, Christopher and Senior, Andrew W. and Shearn, Michael J. and Shorter, Aaron and Shutty, Noah and Shvarts, Vladimir and Singh, Shraddha and Sivak, Volodymyr and Skruzny, Jindra and Small, Spencer and Smelyanskiy, Vadim and Smith, W. Clarke and Somma, Rolando D. and Springer, Sofia and Sterling, George and Strain, Doug and Suchard, Jordan and Szasz, Aaron and Sztein, Alex and Thor, Douglas and Torres, Alfredo and Torunbalci, M. Mert and Vaishnav, Abeer and Vargas, Justin and Vdovichev, Sergey and Vidal, Guifre and Villalonga, Benjamin and Heidweiller, Catherine Vollgraff and Waltman, Steven and Wang, Shannon X. and Ware, Brayden and Weber, Kate and Weidel, Travis and White, Theodore and Wong, Kristi and Woo, Bryan W. K. and Xing, Cheng and Yao, Z. Jamie and Yeh, Ping and Ying, Bicheng and Yoo, Juhwan and Yosri, Noureldin and Young, Grayson and Zalcman, Adam and Zhang, Yaxing and Zhu, Ningfeng and Zobrist, Nicholas},
   year={2024},
   month=dec, pages={920–926} }

@article{Nakamura_1999,
   title={Coherent control of macroscopic quantum states in a single-Cooper-pair box},
   volume={398},
   ISSN={1476-4687},
   number={6730},
   journal={Nature},
   publisher={Springer Science and Business Media LLC},
   author={Nakamura, Y. and Pashkin, Yu. A. and Tsai, J. S.},
   year={1999},
   month=apr, pages={786–788} }

@article{Dennis_2002,
   title={Topological quantum memory},
   volume={43},
   ISSN={1089-7658},
   number={9},
   journal={Journal of Mathematical Physics},
   publisher={AIP Publishing},
   author={Dennis, Eric and Kitaev, Alexei and Landahl, Andrew and Preskill, John},
   year={2002},
   month=sep, pages={4452–4505} }

@inproceedings{mazurek2025tailoredquantumdevicecalibration,
  title={Tailored Quantum Device Calibration with Statistical Model Checking},
  author={Mazurek, Filip and D'Onofrio, Marissa and Van Horn, Andrew and Yu, Jiyong and Ranawat, Kavyashree and Kim, Jungsang and Brown, Kenneth R},
  booktitle={2025 IEEE International Conference on Quantum Computing and Engineering (QCE)},
  volume={1},
  pages={394--404},
  year={2025},
  organization={IEEE}
}

@book{Nielsen_Chuang_2010, 
    place={Cambridge}, 
    title={Quantum Computation and Quantum Information: 10th Anniversary Edition}, 
    publisher={Cambridge University Press}, 
    author={Nielsen, Michael A. and Chuang, Isaac L.}, 
    year={2010}
}

@article{48651,title = {Quantum Supremacy using a Programmable Superconducting Processor},author	= {Frank Arute and Kunal Arya and Ryan Babbush and Dave Bacon and Joseph Bardin and Rami Barends and Rupak Biswas and Sergio Boixo and Fernando Brandao and David Buell and Brian Burkett and Yu Chen and Jimmy Chen and Ben Chiaro and Roberto Collins and William Courtney and Andrew Dunsworth and Edward Farhi and Brooks Foxen and Austin Fowler and Craig Michael Gidney and Marissa Giustina and Rob Graff and Keith Guerin and Steve Habegger and Matthew Harrigan and Michael Hartmann and Alan Ho and Markus Rudolf Hoffmann and Trent Huang and Travis Humble and Sergei Isakov and Evan Jeffrey and Zhang Jiang and Dvir Kafri and Kostyantyn Kechedzhi and Julian Kelly and Paul Klimov and Sergey Knysh and Alexander Korotkov and Fedor Kostritsa and Dave Landhuis and Mike Lindmark and Erik Lucero and Dmitry Lyakh and Salvatore Mandrà and Jarrod Ryan McClean and Matthew McEwen and Anthony Megrant and Xiao Mi and Kristel Michielsen and Masoud Mohseni and Josh Mutus and Ofer Naaman and Matthew Neeley and Charles Neill and Murphy Yuezhen Niu and Eric Ostby and Andre Petukhov and John Platt and Chris Quintana and Eleanor G. Rieffel and Pedram Roushan and Nicholas Rubin and Daniel Sank and Kevin J. Satzinger and Vadim Smelyanskiy and Kevin Jeffery Sung and Matt Trevithick and Amit Vainsencher and Benjamin Villalonga and Ted White and Z. Jamie Yao and Ping Yeh and Adam Zalcman and Hartmut Neven and John Martinis},year	= {2019},journal	= {Nature},pages	= {505–510},volume	= {574}}

@article{Knill_2008,
   title={Randomized benchmarking of quantum gates},
   volume={77},
   ISSN={1094-1622},
   number={1},
   journal={Physical Review A},
   publisher={American Physical Society (APS)},
   author={Knill, E. and Leibfried, D. and Reichle, R. and Britton, J. and Blakestad, R. B. and Jost, J. D. and Langer, C. and Ozeri, R. and Seidelin, S. and Wineland, D. J.},
   year={2008},
   month=jan }

@article{Wu_2025,
   title={Mitigating cosmic-ray-like correlated events with a modular quantum processor},
   volume={24},
   ISSN={2331-7019},
   number={4},
   journal={Physical Review Applied},
   publisher={American Physical Society (APS)},
   author={Wu, Xuntao and Joshi, Yash J. and Yan, Haoxiong and Andersson, Gustav and Anferov, Alexander and Conner, Christopher R. and Karimi, Bayan and King, Amber M. and Li, Shiheng and Malc, Howard L. and Miller, Jacob M. and Mishra, Harsh and Qiao, Hong and Ryu, Minseok and Xing, Siyuan and Shi, Jian and Cleland, Andrew N.},
   year={2025},
   month=oct }

@article{preskill1997faulttolerantquantumcomputation,
  title={Fault-tolerant quantum computation},
  author={Preskill, John},
  journal={Introduction to quantum computation and information},
  volume={213},
  year={1998},
  publisher={World Scientific}
}

@article{Qing:2024xfv,
    author = "Qing, Bingcheng and others",
    title = "{Quantum benchmarking of high-fidelity noise-biased operations on a detuned Kerr-cat qubit}",
    eprint = "2411.04442",
    archivePrefix = "arXiv",
    primaryClass = "quant-ph",
    journal = "Proc. Nat. Acad. Sci.",
    volume = "123",
    number = "5",
    pages = "e2520479123",
    year = "2026"
}

@article{Sarovar_2020,
   title={Detecting crosstalk errors in quantum information processors},
   volume={4},
   ISSN={2521-327X},
   journal={Quantum},
   publisher={Verein zur Forderung des Open Access Publizierens in den Quantenwissenschaften},
   author={Sarovar, Mohan and Proctor, Timothy and Rudinger, Kenneth and Young, Kevin and Nielsen, Erik and Blume-Kohout, Robin},
   year={2020},
   month=sept, pages={321} }

@inproceedings{Murali_2020, series={ASPLOS ’20},
   title={Software Mitigation of Crosstalk on Noisy Intermediate-Scale Quantum Computers},
   booktitle={Proceedings of the Twenty-Fifth International Conference on Architectural Support for Programming Languages and Operating Systems},
   publisher={ACM},
   author={Murali, Prakash and Mckay, David C. and Martonosi, Margaret and Javadi-Abhari, Ali},
   year={2020},
   month=mar, pages={1001–1016},
   collection={ASPLOS ’20} }

@INPROCEEDINGS{11250000,
  author={Safi, Hila and Niedermeier, Christoph and Mauerer, Wolfgang},
  booktitle={2025 IEEE International Conference on Quantum Computing and Engineering (QCE)}, 
  title={Twiddle: Twirling and Dynamical Decoupling, and Crosstalk Noise Modeling}, 
  year={2025},
  volume={02},
  number={},
  pages={162-168} }

@article{Wagner_2025,
   title={Optimized Noise Suppression for Quantum Circuits},
   volume={37},
   ISSN={1526-5528},
   number={1},
   journal={INFORMS Journal on Computing},
   publisher={Institute for Operations Research and the Management Sciences (INFORMS)},
   author={Wagner, Friedrich and Egger, Daniel J. and Liers, Frauke},
   year={2025},
   month=jan, pages={22–41} }

@inproceedings{niu2021analyzingcrosstalkerrornisq,
  title={Analyzing crosstalk error in the NISQ era},
  author={Niu, Siyuan and Todri-Sanial, Aida},
  booktitle={2021 IEEE Computer Society Annual Symposium on VLSI (ISVLSI)},
  pages={428--430},
  year={2021},
  organization={IEEE}
}

@article{PhysRevA.86.062318,
  title = {Surface code with decoherence: An analysis of three superconducting architectures},
  author = {Ghosh, Joydip and Fowler, Austin G. and Geller, Michael R.},
  journal = {Phys. Rev. A},
  volume = {86},
  issue = {6},
  pages = {062318},
  numpages = {12},
  year = {2012},
  month = {Dec},
  publisher = {American Physical Society},
}

@article{Geller_2013,
   title={Efficient error models for fault-tolerant architectures and the Pauli twirling approximation},
   volume={88},
   ISSN={1094-1622},
   number={1},
   journal={Physical Review A},
   publisher={American Physical Society (APS)},
   author={Geller, Michael R. and Zhou, Zhongyuan},
   year={2013},
   month=july }

@article{Burnett_2019,
   title={Decoherence benchmarking of superconducting qubits},
   volume={5},
   ISSN={2056-6387},
   number={1},
   journal={npj Quantum Information},
   publisher={Springer Science and Business Media LLC},
   author={Burnett, Jonathan J. and Bengtsson, Andreas and Scigliuzzo, Marco and Niepce, David and Kudra, Marina and Delsing, Per and Bylander, Jonas},
   year={2019},
   month=june }

@article{PhysRevX.9.041041,
  title = {High-Coherence Fluxonium Qubit},
  author = {Nguyen, Long B. and Lin, Yen-Hsiang and Somoroff, Aaron and Mencia, Raymond and Grabon, Nicholas and Manucharyan, Vladimir E.},
  journal = {Phys. Rev. X},
  volume = {9},
  issue = {4},
  pages = {041041},
  numpages = {14},
  year = {2019},
  month = {Nov},
  publisher = {American Physical Society},
}

@article{PhysRevLett.130.267001,
  title = {Millisecond Coherence in a Superconducting Qubit},
  author = {Somoroff, Aaron and Ficheux, Quentin and Mencia, Raymond A. and Xiong, Haonan and Kuzmin, Roman and Manucharyan, Vladimir E.},
  journal = {Phys. Rev. Lett.},
  volume = {130},
  issue = {26},
  pages = {267001},
  numpages = {6},
  year = {2023},
  month = {Jun},
  publisher = {American Physical Society},
}

@article{Trappen_2025,
   title={Decoherence of a tunable capacitively shunted flux qubit},
   volume={8},
   ISSN={2399-3650},
   number={1},
   journal={Communications Physics},
   publisher={Springer Science and Business Media LLC},
   author={Trappen, Robbyn and Dai, Xi and Yurtalan, M. Ali and Melanson, Denis and Tennant, Daniel M. and Martinez, Antonio J. and Tang, Yongchao and Gibson, Joseph and Grover, Jeffrey A. and Disseler, Steven M. and Basham, James I. and Das, Rabindra and Kim, David K. and Melville, Alexander J. and Niedzielski, Bethany M. and Hirjibehedin, Cyrus F. and Serniak, Kyle and Weber, Steven J. and Yoder, Jonilyn L. and Oliver, William D. and Lidar, Daniel A. and Lupascu, Adrian},
   year={2025},
   month=nov }

@article{Kawakami_2016,
   title={Gate fidelity and coherence of an electron spin in an Si/SiGe quantum dot with micromagnet},
   volume={113},
   ISSN={1091-6490},
   number={42},
   journal={Proceedings of the National Academy of Sciences},
   publisher={Proceedings of the National Academy of Sciences},
   author={Kawakami, Erika and Jullien, Thibaut and Scarlino, Pasquale and Ward, Daniel R. and Savage, Donald E. and Lagally, Max G. and Dobrovitski, Viatcheslav V. and Friesen, Mark and Coppersmith, Susan N. and Eriksson, Mark A. and Vandersypen, Lieven M. K.},
   year={2016},
   month=oct, pages={11738–11743} }

@article{Stano_2022,
   title={Review of performance metrics of spin qubits in gated semiconducting nanostructures},
   volume={4},
   ISSN={2522-5820},
   number={10},
   journal={Nature Reviews Physics},
   publisher={Springer Science and Business Media LLC},
   author={Stano, Peter and Loss, Daniel},
   year={2022},
   month=aug, pages={672–688} }

@article{RevModPhys.95.025003,
  title = {Semiconductor spin qubits},
  author = {Burkard, Guido and Ladd, Thaddeus D. and Pan, Andrew and Nichol, John M. and Petta, Jason R.},
  journal = {Rev. Mod. Phys.},
  volume = {95},
  issue = {2},
  pages = {025003},
  numpages = {58},
  year = {2023},
  month = {Jun},
  publisher = {American Physical Society},
}

@misc{IBMQuantum,
  author = {{IBM Quantum}},
  title  = {{IBM Quantum}},
  year   = {2026},
  url    = {https://quantum.cloud.ibm.com/computers?processorType=Heron}
}

@article{PhysRevA.107.052417,
  title = {Improving trapped-ion-qubit memories via code-mediated error-channel balancing},
  author = {Seis, Yannick and Brown, Benjamin J. and S\o{}rensen, Anders S. and Goodwin, Joseph F.},
  journal = {Phys. Rev. A},
  volume = {107},
  issue = {5},
  pages = {052417},
  numpages = {15},
  year = {2023},
  month = {May},
  publisher = {American Physical Society},
}

@article{PhysRevA.109.032433,
  title = {Tailoring quantum error correction to spin qubits},
  author = {Het\'enyi, Bence and Wootton, James R.},
  journal = {Phys. Rev. A},
  volume = {109},
  issue = {3},
  pages = {032433},
  numpages = {22},
  year = {2024},
  month = {Mar},
  publisher = {American Physical Society},
}

@article{PhysRevApplied.20.044045,
  title = {Long spin coherence and relaxation times in nanodiamonds milled from polycrystalline ${}^{12}$$\mathrm{C}$ diamond},
  author = {March, James E. and Wood, Benjamin D. and Stephen, Colin J. and Fervenza, Laura Dur\'an and Breeze, Ben G. and Mandal, Soumen and Edmonds, Andrew M. and Twitchen, Daniel J. and Markham, Matthew L. and Williams, Oliver A. and Morley, Gavin W.},
  journal = {Phys. Rev. Appl.},
  volume = {20},
  issue = {4},
  pages = {044045},
  numpages = {11},
  year = {2023},
  month = {Oct},
  publisher = {American Physical Society},
}

@article{PRXQuantum.4.020350,
  title = {One Hundred Second Bit-Flip Time in a Two-Photon Dissipative Oscillator},
  author = {Berdou, C. and Murani, A. and R\'eglade, U. and Smith, W.C. and Villiers, M. and Palomo, J. and Rosticher, M. and Denis, A. and Morfin, P. and Delbecq, M. and Kontos, T. and Pankratova, N. and Rautschke, F. and Peronnin, T. and Sellem, L.-A. and Rouchon, P. and Sarlette, A. and Mirrahimi, M. and Campagne-Ibarcq, P. and Jezouin, S. and Lescanne, R. and Leghtas, Z.},
  journal = {PRX Quantum},
  volume = {4},
  issue = {2},
  pages = {020350},
  numpages = {20},
  year = {2023},
month = {Jun},
  publisher = {American Physical Society},
}

@article{PhysRevA.106.062428,
  title = {Performance of surface codes in realistic quantum hardware},
  author = {iOlius, Antonio deMarti and Martinez, Josu Etxezarreta and Fuentes, Patricio and Crespo, Pedro M. and Garcia-Frias, Javier},
  journal = {Phys. Rev. A},
  volume = {106},
  issue = {6},
  pages = {062428},
  numpages = {18},
  year = {2022},
  month = {Dec},
  publisher = {American Physical Society},
}

@article{Preskill_2018,
   title={Quantum Computing in the NISQ era and beyond},
   volume={2},
   ISSN={2521-327X},
   journal={Quantum},
   publisher={Verein zur Forderung des Open Access Publizierens in den Quantenwissenschaften},
   author={Preskill, John},
   year={2018},
   month=Aug, pages={79} }
\end{document}